\documentclass[a4paper,11pt]{article}
\usepackage{pos}
\usepackage{physics}
\usepackage{placeins}
\usepackage{mathrsfs}
\usepackage{hyperref}

\title{Controlling unwanted exponentials in lattice calculations of radiative leptonic decays}

\author*[a]{Christopher Kane}
\author[b]{Davide Giusti}
\author[b,c]{Christoph Lehner}
\author[a]{Stefan Meinel}
\author[c]{Amarjit Soni}

\affiliation[a]{Department of Physics, University of Arizona\\
  Tucson, AZ 85719, USA}

\affiliation[b]{Department of Physics, University of Regensburg, 93040 Regensburg, Germany}

\affiliation[c]{Physics Department, Brookhaven National Laboratory, Upton, NY 11973, USA}

\emailAdd{cfkane@email.arizona.edu}

\abstract{Two important sources of systematic errors in lattice QCD calculations of radiative leptonic decays are unwanted exponentials in the sum over intermediate states and unwanted excited states created by the meson interpolating field. Performing the calculation using a 3d sequential propagator allows for better control over the systematic uncertainties from intermediate states, while using a 4d sequential propagator allows for better control over the systematic uncertainties from excited states. We calculate form factors using both methods and compare how reliably each controls these systematic errors. We also employ a hybrid approach involving global fits to data from both methods.}

\FullConference{%
 The 38th International Symposium on Lattice Field Theory, LATTICE2021
  26th-30th July, 2021
  Zoom/Gather@Massachusetts Institute of Technology
}

\begin{document}
\maketitle

\section{Introduction}
Radiative leptonic decays have been gaining interest in recent years \cite{Beneke:1999br, Korchemsky:1999qb, Descotes-Genon:2002crx, Lunghi:2002ju, Braun:2012kp, Wang:2018wfj, Beneke:2018wjp, Belle:2018jqd, Janowski:2021yvz, Carvunis:2021jga, Khodjamirian:2020hob, Beneke:2021rjf, Beneke:2020fot, Shen:2018abs, Lu:2021ttf, Shen:2020hsp, Albrecht:2019zul, Abbas:2018xdu, Pullin:2021ebn}. One important part of understanding these decay processes is a first principles calculation of the relevant hadronic form factors using lattice QCD. The calculation of the form factors requires calculating a non-local matrix element, that, on the lattice, can be calculated using two different methods, which we call the 3d-method and the 4d-method (see section \ref{section:sequential_propagators}). At Lattice 2019 \cite{Kane:2019jtj}, we presented calculations using the 3d-method in the rest frame of the initial-state pseudoscalar meson and fit the data to a constant where it had plateaued. Since then, a lattice calculation of radiative leptonic decays was published in \cite{Desiderio:2020oej}, where the authors use what we call the 4d-method. In their analysis, they fit the data to constants where it had plateaued. 

In the following, we give an update on our work. To explore a wider range of photon energies, we performed new calculations using the 3d-method in the moving frame of the pseudoscalar and found that for some data fitting to a constant is not possible, and more complicated fits have to be implemented to remove unwanted exponentials. The focus of this work is to determine which method results in the best control of systematic uncertainties from these unwanted exponentials with the smallest statistical uncertainties. We present fit results for both 3d and 4d method data, as well as a hybrid approach where we perform global fits to both sets of data.

\section{Hadronic tensor and form factors}
The Minkowski space hadronic tensor for the decay process $H \to \gamma \ell \nu$, where $H$ is a pseudoscalar meson, is defined as
\begin{equation}
    T_{\mu \nu} = -i \int \dd^4 x \ e^{i p_\gamma \vdot x} \bra{0} \textbf{T} \big( J^\text{em}_\mu(x) J^{\text{weak}}_\nu(0) \big) \ket{H(\va{p}_H)}.
\end{equation}
The electromagnetic current and weak current are given by $J^\text{em}_\mu = \sum_q Q_q \bar{q} \gamma_\mu q$ and $J^{\text{weak}}_\nu = \bar{q}_1\gamma_\nu(1-\gamma_5) q_2 $. For real photons, i.e. $p_\gamma^2=0$, which we focus on in this work, the hadronic tensor can be decomposed as \cite{Beneke:2018wjp}
\begin{equation}
    T_{\mu \nu} = \epsilon_{\mu \nu \tau \rho} p^{\tau}_\gamma v^{\rho} F_V + i\big[ -g_{\mu \nu}(v \vdot p_\gamma) + v_\mu(p_\gamma)_\nu \big] F_A - i \frac{v_\mu v_\nu}{(v \vdot p_\gamma)} m_H f_H + (p_\gamma)_\mu\text{-terms},
\end{equation}
where $p_H^\mu = m_H v^\mu$. The $(p_\gamma)_\mu \text{-terms}$ are proportional to the photon momentum and are zero when contracted with the photon polarization vector. The vector form factor $F_V$ and axial form factor $F_A$ are functions of the photon energy as seen in the rest frame of the pseudoscalar meson, given by $E_\gamma^{(0)} = v \vdot p_\gamma$. We define $x_\gamma = 2E_\gamma^{(0)}/m_H$, which, for physically allowed values of $E_\gamma^{(0)}$, takes on values $0 < x_\gamma \leq 1$. The axial form factor is composed of a point-like contribution, where the photon does not probe the internal structure of the initial state pseudoscalar meson, and a structure-dependent contribution. The point-like contribution is given by $(-Q_\ell \frac{f_H}{E_\gamma^{(0)}})$ where $Q_\ell$ is the charge of the lepton in the final state, and $f_H$ is the pseudoscalar decay constant. The structure-dependent part of the axial form factor is given by $F_{A,SD} = F_A - (-Q_\ell \frac{f_H}{E_\gamma^{(0)}})$. At large photon energies, the decay amplitude depends only on $F_V$ and $F_{A,SD}$ \cite{Beneke:2018wjp}. 

In the next section, we show how to extract the hadronic tensor using a Euclidean three-point function. To do so, we need to look at the spectral decomposition for both the $t_{em}<0$ and $t_{em}>0$ time orderings of $T_{\mu \nu}$. After inserting a complete set of energy/momentum eigenstates and performing time integrals we find
\begin{equation}
	\begin{split}
		T^<_{\mu \nu} &= -i \int_{-\infty(1-i\epsilon)}^0 dt_{em} \int d^3x \ e^{-i p_\gamma \vdot x} \bra{0} J_\nu^{\text{weak}}(0) J^\text{em}_\mu(t_{em}, \va{x}) \ket{H(\va{p}_H)}
		\\
		&= -\sum_{n} \frac{\bra{0} J_\nu^{\text{weak}}(0) \ket{n(\va{p}_H-\va{p}_\gamma)} \bra{n(\va{p}_H-\va{p}_\gamma)} J^\text{em}_\mu(0) \ket{H(\va{p}_H)}}{2 E_{n,\va{p}_H-\va{p}_\gamma}(E_\gamma + E_{n,\va{p}_H-\va{p}_\gamma} - E_{H,\va{p}_H} - i \epsilon)},
	\end{split}
	\label{eq:hadronic_tensor_spectral_lessthan}
\end{equation}
and
\begin{equation}
	\begin{split}
		T^>_{\mu \nu} &= -i \int_{0}^{\infty(1-i\epsilon)} dt_{em} \int d^3x \ e^{-i p_\gamma \vdot x} \bra{0} J^\text{em}_\mu(t_{em}, \va{x}) J_\nu^{\text{weak}}(0) \ket{H(\va{p}_H)}
		\\
		&= -\sum_{m} \frac{\bra{0} J^\text{em}_\mu(0) \ket{m(\va{p}_\gamma)} \bra{m(\va{p}_\gamma)} J_\nu^\text{weak}(0) \ket{H(\va{p}_H)}}{2 E_{m,\va{p}_\gamma}(E_\gamma - E_{m,\va{p}_\gamma} - i \epsilon)},
	\end{split}
	\label{eq:hadronic_tensor_spectral_greaterthan}
\end{equation}
where in infinite volume, the sums over $n$ and $m$ include integrals over the continuous spectrum of multi-particle states.

\section{Extracting the hadronic tensor from a Euclidean three-point function}
The Euclidean-time three-point function we will use to extract $T_{\mu \nu}$ is given by
\begin{equation}
	C_{3, \mu \nu}(t_{em}, t_H) = \int d^3x \int d^3y \ e^{-i \va{p}_\gamma \vdot \va{x}} e^{i \va{p}_H \vdot \va{y}}  \langle J_{\mu}^\text{em}(t_{em}, \va{x}) J_\nu^\text{weak}(0) \phi^\dagger_H(t_H, \va{y}) \rangle,
	\label{eq:three_point}
\end{equation}
where $\phi_H^\dagger = -\bar{q}_2 \gamma_5 q_1$ is our meson interpolating field. We omit the momentum arguments for brevity. Additionally, we define the time-integrated correlation functions for each time ordering,
\begin{equation}
    I^<_{\mu \nu}(t_H, T) = \int^0_{-T} dt_{em} e^{E_\gamma t_{em}} C_{3, \mu \nu}(t_{em}, t_H), \hspace{0.1in} I^>_{\mu \nu}(t_H, T) = \int^T_{0} dt_{em} e^{E_\gamma t_{em}} C_{3, \mu \nu}(t_{em}, t_H),
\end{equation}
for a finite integration range $T$. Inserting complete sets of energy/momentum eigenstates in our three-point correlation function and performing the Euclidean time integrals we find,
\begin{align}
    \begin{split}\label{eq:Imunu_spectral_lessthan}
        I^<_{\mu \nu}(t_H, T) &= \sum_{l,n} \frac{\bra{0} J^\text{weak}_\nu(0) \ket{n(\va{p}_H - \va{p}_\gamma)} \bra{n(\va{p}_H - \va{p}_\gamma)} J^\text{em}_\mu(0) \ket{l(\va{p}_H)} \bra{l(\va{p}_H)} \phi^\dagger_H(0) \ket{0}}{2E_{n, \va{p}_H - \va{p}_\gamma} 2E_{l, \va{p}_H}(E_\gamma + E_{n,\va{p}_H - \va{p}_\gamma} - E_{l,\va{p}_H})}
       \\
       &\hspace{0.25in} \times e^{E_{l, \va{p}_H} t_H}\Big[1 - e^{-(E_\gamma - E_{l, \va{p}_H} + E_{n, \va{p}_H-\va{p}_\gamma})T}\Big],
    \end{split}
    \\
    \begin{split}\label{eq:Imunu_spectral_greaterthan}
        I^>_{\mu \nu}(t_H, T) &= \sum_{l,m} \frac{\bra{0} J^\text{em}_\mu(0) \ket{m(\va{p}_\gamma)} \bra{m(\va{p}_\gamma)}  J^\text{weak}_\nu(0)  \ket{l(\va{p}_H)} \bra{l(\va{p}_H)}  \phi^\dagger_H(0) \ket{0}}{2E_{m, \va{p}_\gamma}2E_{l, \va{p}_H}(E_\gamma - E_{m,\va{p}_\gamma})} 
        \\
        &\hspace{0.25in} \times e^{E_{l,\va{p}_H} t_H} \big[ e^{(E_\gamma - E_{m, \va{p}_\gamma})T} - 1 \big].
    \end{split}
\end{align}
Taking the limit $t_H \to -\infty$ removes excited state contamination from the interpolating field $\phi^\dagger_H$. We see that the time-integrated correlation function contains the sum over all desired intermediate states, but because of the finite integration range $T$, each state comes with an unwanted exponential. In \cite{Kane:2019jtj} we argued that as long as $|\va{p}_\gamma| > 0$, the unwanted exponentials for both time orderings decay as we increase the integration range $T$ and thus we have the final relation
\begin{equation}
	T_{\mu \nu} = - \lim_{T \to \infty} \lim_{t_H \to -\infty} \frac{2 E_H(\va{p}_H) e^{-E_H(\va{p}_H) t_H}}{\bra{H(\va{p}_H)} \phi^\dagger_H(0) \ket{0}} I_{\mu \nu}(t_H, T),
	\label{eq:Tmunu_from_correlation}
\end{equation}
where $I_{\mu \nu}(t_H, T) = I^<_{\mu \nu}(t_H, T)+I^>_{\mu \nu}(t_H, T)$. It will be useful for our discussion in section \ref{section:fit_method} to introduce the notation $I_{\mu \nu}(t_H,T) = I^A_{\mu \nu}(t_H, T) + I^V_{\mu \nu}(t_H, T)$, where $I^A_{\mu \nu}(t_H, T)$ and $I^V_{\mu \nu}(t_H,T)$ are the weak axial-vector and vector current components of $I_{\mu \nu}(t_H,T)$, respectively.

\section{Sequential propagators} \label{section:sequential_propagators}
\begin{figure}[h]
	\centering
	\begin{minipage}{0.45\textwidth}
	\centering
		\includegraphics[width=0.8\textwidth]{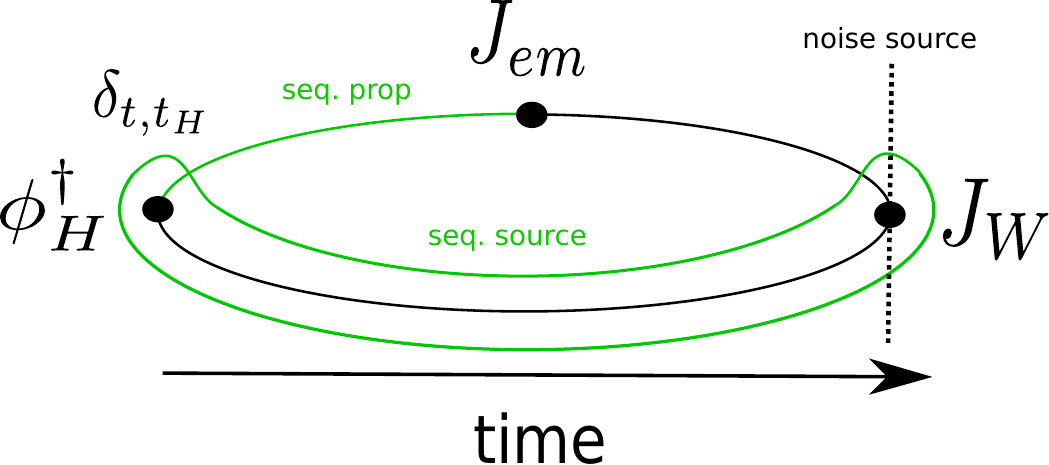}	
	\end{minipage}
	\begin{minipage}{0.45\textwidth}
	\centering
		\includegraphics[width=0.8\textwidth]{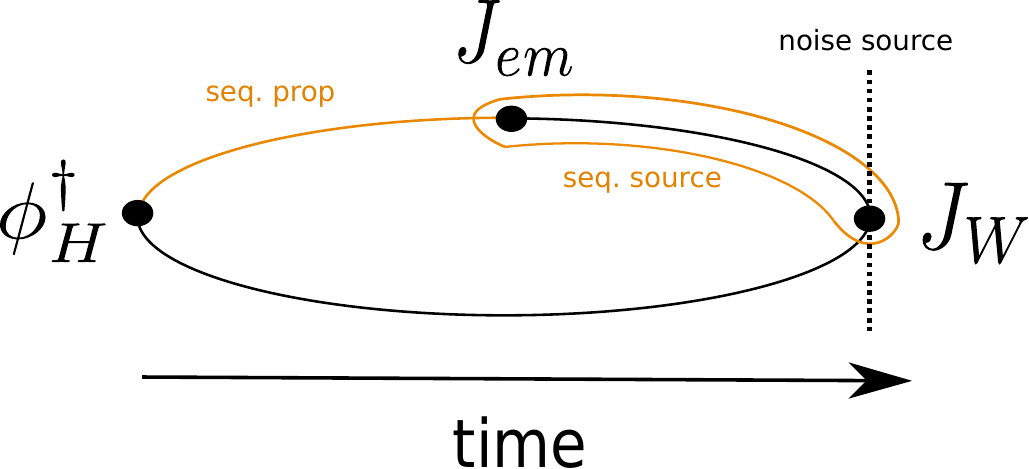}	
	\end{minipage}
	\caption{The left (right) figure is a schematic drawing of the 3d (4d) methods. For both methods, the initial noise source is located at the weak current time. The sequential propagator is shown in green (orange) and the sequential source is circled in green (orange).}
	\label{fig:seq_prop_quark_lines}	
\end{figure}

We consider two different methods of calculating $I_{\mu \nu}(t_H, T)$ on the lattice, which are depicted in figure \ref{fig:seq_prop_quark_lines}. The first, which we call the 3d method, uses a 3d (timeslice) sequential propagator through the interpolating field. Using the 3d method, for a fixed value of $t_H$ we calculate the three-point correlation function in equation (\ref{eq:three_point}) and  get all values of $t_{em}$ for free. The time integral over $t_{em}$ is performed offline in the analysis stage. The other method, which we call the 4d method, uses a 4d sequential propagator through the EM current which is not fixed to a single timeslice. The key difference is that using the 4d method, for a fixed value of integration range $T$, \textit{the time integral over $t_{em}$ is performed directly on the lattice}, such that we get all values of $t_H$ for free. We see that the 3d method is particularly suited to control unwanted exponentials from finite integration range $T$, while the 4d method is particularly suited to control unwanted exponentials from excited states created by the interpolating field. In \cite{Kane:2019jtj}, we used the 3d method and performed fits to a constant where the data plateaued in $t_H$ and $T$. The results in \cite{Desiderio:2020oej} were calculated using the 4d method and integrated over the entire time extent of the lattice, i.e. $T=N_T/2$. 

In this work, we performed calculations using the 3d method for multiple values of $t_H$ and the 4d method for multiple values of $T$.
\vspace{-0.1in}
\section{Lattice parameters}
We perform calculations on two RBC/UKQCD ensembles, namely the ``24I'' ensemble \cite{RBC:2010qam} and a $32^3 \times 64$ ensemble with identical properties as the ``24I'', but with a larger spatial volume. Both ensembles were generated using the Iwasaki gauge action, 2+1 flavors of domain-wall fermions with $N_5=16$ sites in the fifth dimension, $\beta=2.13, am_{u,d}=0.005, am_s^\text{sea}=0.04,$ and have $a^{-1}=1.785(5) \text{ GeV}$. For the light and strange quarks we use the same domain-wall action as the sea quarks in \cite{RBC:2010qam}, except that for the strange quark we use the physical mass of $am_s^\text{val}=0.0323$ rather than the sea mass. We implement the valence charm quark using a M\"{o}bius domain-wall action with stout-smeared gauge links (N=3, $\rho=0.1$), $L_5/a=12$, $aM_5=1.0$, $am_f=0.6$ \cite{Boyle:2018knm}, which approximately corresponds to the physical charm-quark mass. Disconnected diagrams are currently neglected. In our calculation, we use all-mode averaging \cite{Shintani:2014vja} with 1 exact and 16 sloppy samples per configuration, where the sloppy samples correspond to 16 different starting time slices for the noise source. We also use local currents with ``mostly non-perturbative'' renormalization. For all 3d-method data we performed calculations for three values of source-sink separation $-t_H/a=\{6,9,12\}$. For all 4d-method data we performed calculations for three values of integration range $T/a=\{6,9,12\}$. Further details of the calculation are shown in Table \ref{tab:parameter_details}.
%
\begin{table}[h]
\centering
\begin{tabular}{c|c|c|c|c|c}
	\hline
	\hline
	Meson & $N_s^3 \times N_t$ & $N_{\text{cfg}}$ & Method & & \\
	\hline
	$K$ & $32^3 \times 64$ & 20 & 3d & $p_{K,z}$ = $2\pi/L \{1, 2\}$ & $p_{\gamma,z} = 2\pi/L \{1\}$ \\
	$K$ & $32^3 \times 64$ & 20 & 4d & $p_{K,z}$ = $2\pi/L \{1, 2\}$ & $p_{\gamma,z} = 2\pi/L \{1\}$ \\
	\hline
	$D_s$ & $24^3 \times 64$ & 25 & 3d & $|\va{p}_{D_s}|=0$ & $|\va{p}_\gamma|^2=(2\pi/L)^2\{1,2,3,4\}$ 
	\\
	$D_s$ & $24^3 \times 64$ & 25 & 4d & $p_{D_s,z}=2\pi/L\{-1,0,1,2\}$ & $p_{\gamma,z}=(2\pi/L)\{1\}$
\end{tabular}
\caption{The number of configurations, methods, and momenta for which we performed calculations. When only the z-component of the momentum is listed, the other momentum components are zero.}
\label{tab:parameter_details}
\end{table}
\vspace{-0.1in}
\section{Fit Method} \label{section:fit_method}
In this section we describe our fit method used to remove unwanted exponentials from the form factors. We begin by considering, in continuum QCD, the quantum numbers of the states that contribute to the sum over states in the spectral decompositions of $I_{\mu \nu}^<(t_H, T)$ and $I_{\mu \nu}^>(t_H, T)$. For $t_{em}<0$, the states must have the same quark-flavor quantum numbers as the initial pseudoscalar meson. Additionally, parity constrains the possible $J^P$ quantum numbers that contribute. The $J^P$ quantum numbers of the states $\ket{n(\va{p}_H-\va{p}_\gamma)}$ that contribute to the sum over states in the spectral decomposition of $I_{\mu \nu}^{<,A}$ are $J^P = \{0^-,1^+,2^\pm, \dots\}$ (on the lattice, the states are in irreducible representations of the associated little group of the cubic group, which mixes angular momentum quantum numbers). The lowest-energy state with these quantum numbers is the pseudoscalar meson itself. For $I_{\mu \nu}^{<,V}$, the states $\ket{n(\va{p}_H-\va{p}_\gamma)}$ that contribute to the sum over all states have $J^P = \{0^+, 1^-, 2^\pm, \dots\}$. The lowest energy state with these quantum numbers is the vector meson ($H^*$) associated with our pseudoscalar meson, e.g. for $H=K$ it would be a ($K^*$)-like state. We calculate the energies of both the $H$ and $H^*$ by fitting the associated two-point function to a single exponential and use the result of the fit as a Gaussian prior in the form factor fits, where the central value of the fit result is the prior value and the uncertainty of the fit result is the prior width. We use the continuum relativistic dispersion relation to calculate energies at non-zero momentum for the $K$ and $K^*$ mesons. For the $D_s$ and $D_s^*$ we calculate non-zero momentum energies directly from the two-point correlation function projected to definite momentum. For $t_{em}>0$, the states are flavorless and we leave their energies as fit parameters. Considering parity, the quantum numbers of the states $\ket{m(\va{p}_\gamma)}$ that contribute to the sum over states in $I_{\mu \nu}^{>,A}(t_H,T)$ and $I_{\mu \nu}^{>,V}(t_H,T)$ are $J^P = \{0^+, 1^-, 2^\pm, \dots\}$ and $J^P = \{1^-, 2^\pm, \dots\}$, respectively. \looseness=-1

From this we also learn that, for a given time ordering, the same states contribute to all $\mu, \nu$ components of $I^A_{\mu \nu}(t_H, T)$, and similarly for $I^V_{\mu \nu}(t_H, T)$. So, while the matrix elements multiplying the unwanted exponentials will in general be different for different $\mu,\nu$, the energies appearing in the unwanted exponentials will be the same. Because only $I^A_{\mu \nu}$ contributes to $F_A, F_{A,SD}, f_H$ and only $I^V_{\mu \nu}$ contributes to $F_V$, we can fit the form factors directly without mixing unwanted exponentials. We choose to fit the form factors instead of $I_{\mu \nu}$ for two reasons. First, fitting the form factors requires fewer total fit parameters which helps stabilize the fits. Second, imagine the scenario where taking linear combinations of $I_{\mu \nu}(t_H, T)$ results in cancellations which reveal features in the form factors that $I_{\mu \nu}(t_H, T)$ is not sensitive to. If we fit $I_{\mu \nu}(t_H, T)$ first, these features could be missed by the fit and propagate as a source of systematic uncertainties to the form factors. Fitting the form factors directly removes this possibility.

To help constrain the energy gap between the first excited state and ground created by the interpolating field, $\Delta E$, we first perform two-exponential fits to the pseudoscalar two-point function and use the fit result for $\Delta E$ as a Gaussian prior in the form factor fits. We extract $F_V, F_A$ and $f_H$ from the time-integrated correlation function. Using the extracted values of $F_A$ and $f_H$, we then calculate the structure dependent axial form factor by $F_{A,SD}=F_A-(-Q_l \frac{f_H}{E_\gamma^{(0)}})$. To take advantage of the fact that data on a given ensemble will have common energies appearing in the unwanted exponentials that come with the intermediate states, as well as the excited state energy gap from the interpolating field, we perform simultaneous fits to all data calculated on a given ensemble.

\begin{figure}
	\centering
	\begin{minipage}{0.47\textwidth}
		\includegraphics[width=0.9\textwidth]{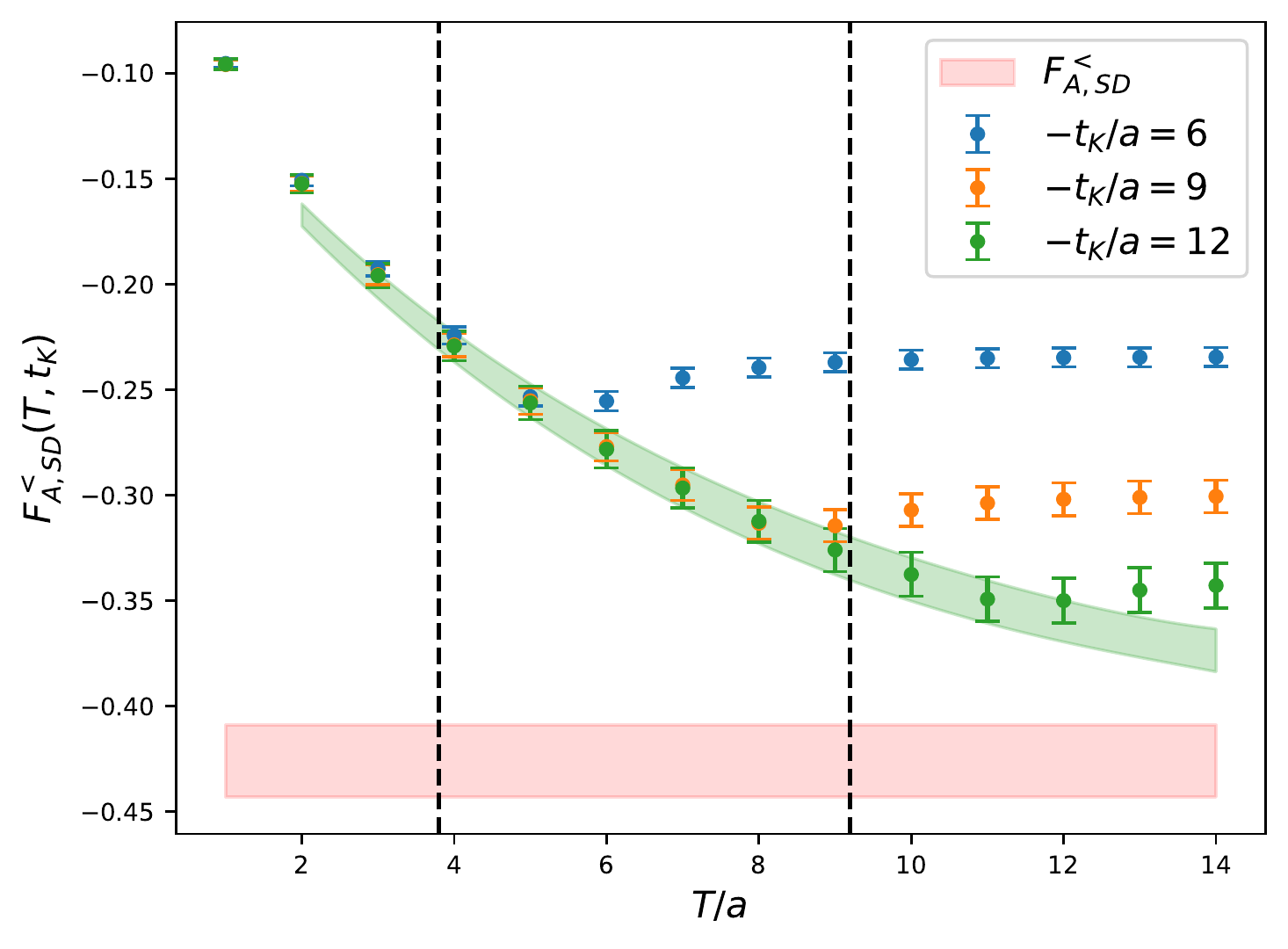}
	\end{minipage}
	\begin{minipage}{0.47\textwidth}
		\includegraphics[width=0.9\textwidth]{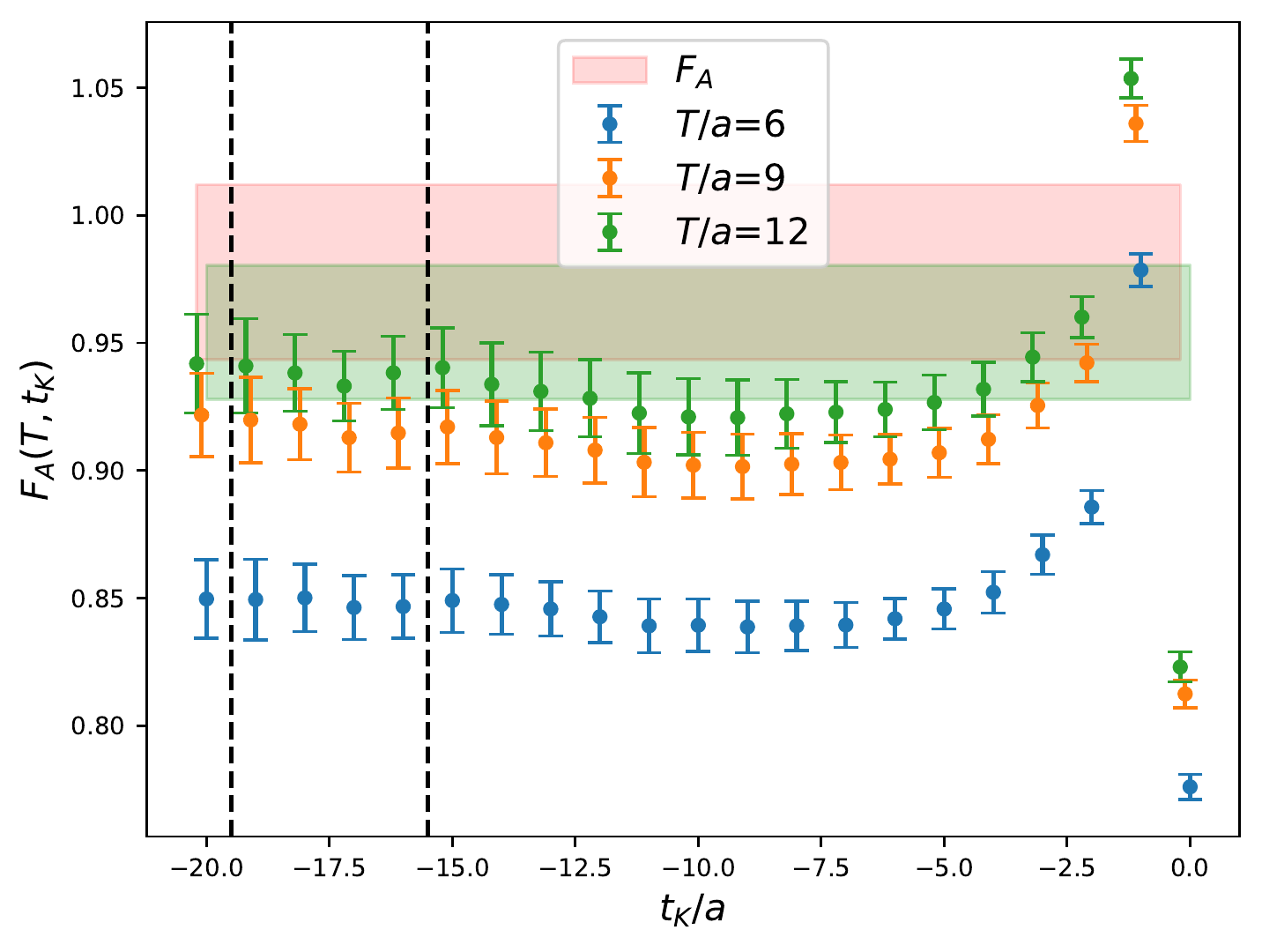}
	\end{minipage}
	\caption{Left: $F^<_{A,SD}(t_K, T)$ calculated using the 3d method as a function of $T$ for three values of $t_K$. The green band is the one sigma fit result for $-t_K/a=12$ and the vertical dotted lines indicate the fit range. The red band is the one sigma extrapolated value of $F^<_{A,SD}$. Right: $F_A(t_K, T)$ calculated using the 4d method as a function of $t_K$ for three values of $T$. The vertical dashed lines indicate the fit range, and the green band is the one sigma fit result for $T/a=12$. The red band is the one sigma fit result for $F_A$. Both were calculated with $\va{p}_K=\frac{2\pi}{L}(0,0,1), \va{p}_\gamma=\frac{2\pi}{L}(0,0,1)$.}
	\label{fig:form_factors_fits_Kaon}
\end{figure}

We are fitting our data as a function of integration range, and so each successive value of $T$ is directly dependent on smaller values of $T$. These large correlations lead to small eigenvalues in the correlation matrix, making correlated fits to this data unstable. This, combined with the fact that our global fits have up to $\order{100}$ fit parameters, means that performing correlated fits is not possible. We therefore perform uncorrelated fits and calculate the central values and statistical uncertainties using jackknife. Before performing the global fits, we first determine stable fit ranges for each form factor at a given momentum on a given ensemble. The stable fit ranges for the 3d method are chosen by performing simultaneous fits to all $t_H$ while looking for stability in the $T$ fit range. For the 4d method we perform simultaneous fits to all values of $T$ looking for stability in the $t_H$ fit range. The chosen stable fit ranges are then used in the global fits.
\begin{figure}[h]
	\centering
	\begin{minipage}{0.46\textwidth}
		\includegraphics[width=0.9\textwidth]{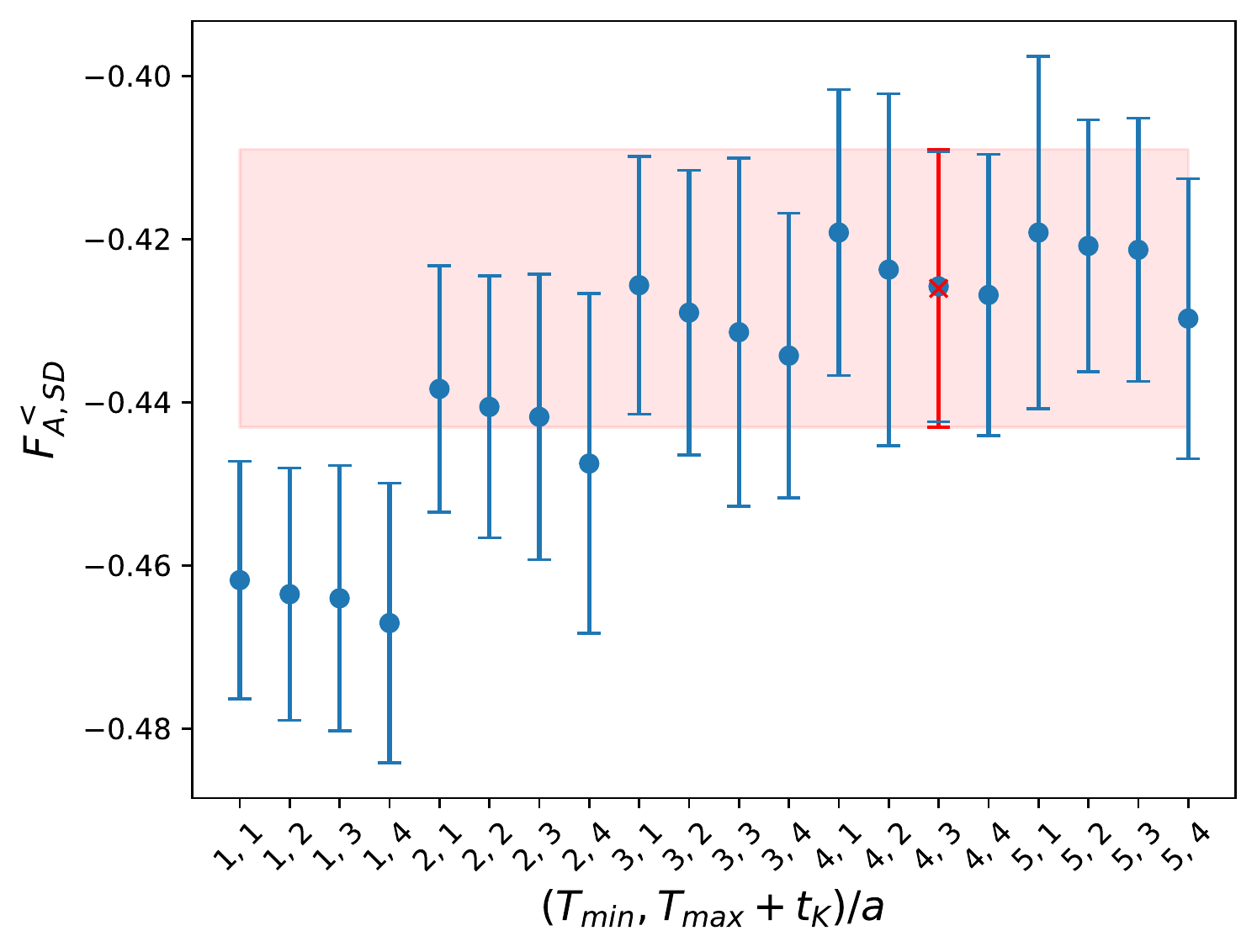}
	\end{minipage}
	\begin{minipage}{0.53\textwidth}
		\includegraphics[width=0.9\textwidth]{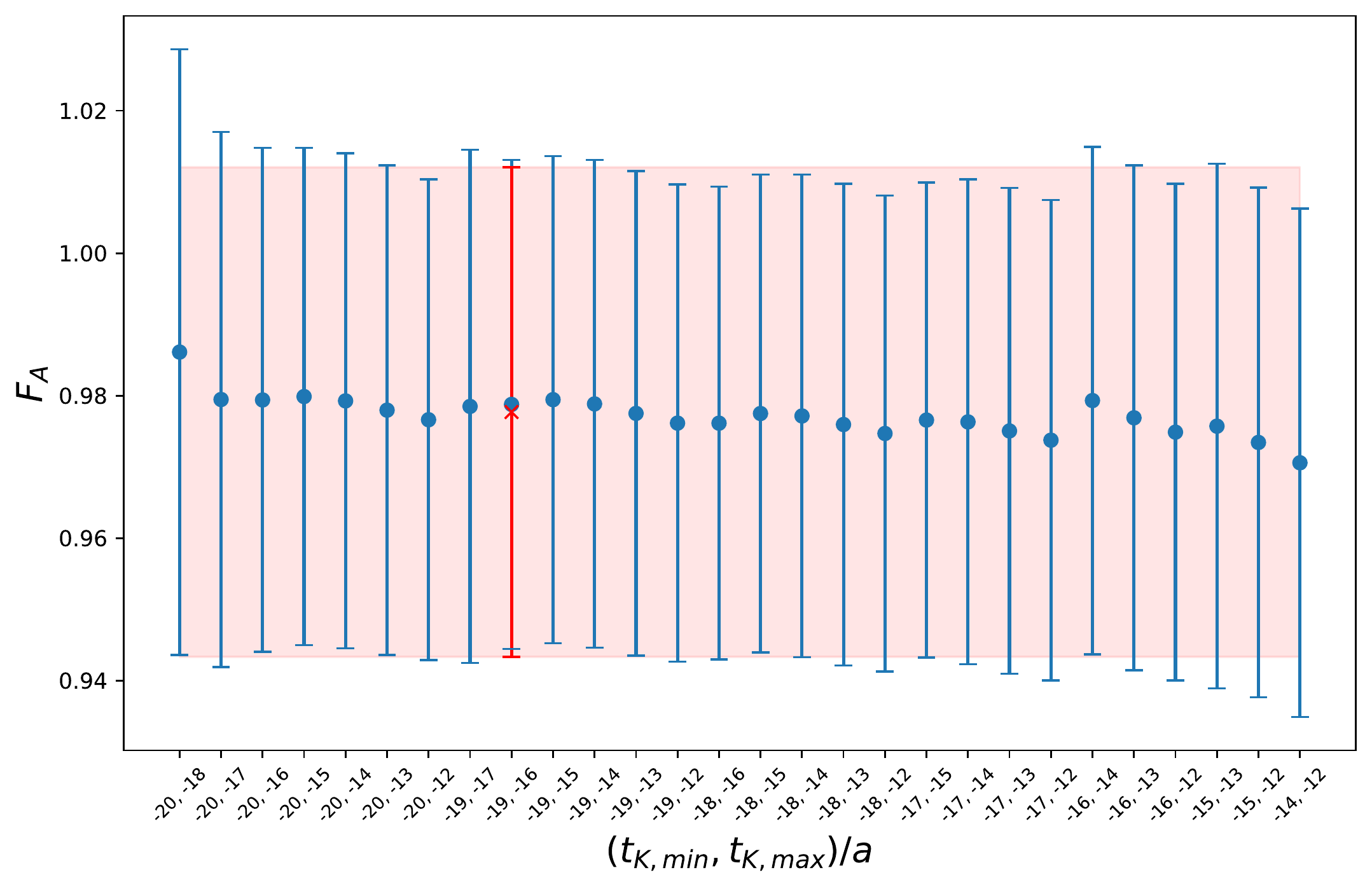}
	\end{minipage}
	\caption{Left: $F^<_{A,SD}$ calculated using the 3d method as a function of fit range $(T_\text{min}, T_\text{max}+t_K)/a$. The red point is the chosen stable fit range and the result of the global fit to all 3d method data. Fit ranges where $-t_K/a=6$ has no data points means it was left out of the fit. Right: $F_A$ calculated using the 4d method as a function of fit range $(t_{K,\text{min}}, t_{K,\text{max}})/a$. Both were calculated with $\va{p}_K=\frac{2\pi}{L}(0,0,1), \va{p}_\gamma=\frac{2\pi}{L}(0,0,1)$.}
	\label{fig:stability_plots_Kaon}
\end{figure}
The fit form for the 3d method data includes one exponential to account for the unwanted exponential that comes with the lowest-energy intermediate state, and one exponential to account for the unwanted exponential from the lowest energy excited state created by the interpolating field. The fit form for the $t_{em}<0$ and $t_{em}>0$ time orderings for a variable $F=F_V, F_{A,SD}, F_A, f_H$ are given by
\begin{align}
    F^{<}(t_H, T) &= F^< + B_{F}^{<} \big(1+ B^{<}_{F, \text{exc}} e^{\Delta E (T+t_H)} \big) e^{-(E_\gamma - E_{H} + E^<)T} + C^{<}_{F} e^{\Delta E t_H},
    \label{eq:3d_fit_form_neg}
    \\
    F^{>}(t_H, T) &= F^> + B_{F}^{>} \big(1+ B^{<}_{F, \text{exc}}e^{\Delta E t_H} \big) e^{(E_\gamma - E^>)T} + C^{>}_{F} e^{\Delta E t_H}.
    \label{eq:3d_fit_form_pos}
\end{align}
Notice that for $t_{em}<0$, for finite $t_H$, one must be careful to not integrate all the way back to the interpolating field, i.e. $T < -t_H$. For $t_{em}<0$, the stability checks are done by looking at the minimum fit range and the distance from the interpolating field. For $t_{em}>0$ we only need to look for stability in the minimum fit range. The 4d data is a sum of both time orderings and the general fit form would be a sum of those in equations (\ref{eq:3d_fit_form_neg}) and (\ref{eq:3d_fit_form_pos}). However, we perform fits to regions where the data has plateaued in $t_H$, leading to the following fit form
\begin{align}
    F(T) &= F + B_{F}^{<} e^{-(E_\gamma - E_{H} + E^<)T} + B_{F}^{>} e^{(E_\gamma - E^>)T}.
    \label{eq:4d_fit_form}
\end{align}
Even though the energy $E^<$ is constrained from the two-point correlation function, because we only have three values of $T$, the fits to 4d data are not stable. To stabilize the fits we put a broad Gaussian prior on the parameter $E^>$. For the Kaon, the prior is centered at the $\rho$ meson mass with a width of 150 MeV, and for the $D_s$, the prior is centered at the $\phi$ meson mass with a width of 200 MeV. Figure \ref{fig:stability_plots_Kaon} shows example Kaon stability fit plots for the 3d and 4d method fits. We find that, in general, the global fit does not significantly reduce the statistical errors. Figure \ref{fig:form_factors_fits_Kaon} shows an example of the global fit function on top of the Kaon data for the 3d and 4d method. Note that all uncertainties in the plots are purely statistical.
\section{Global fit results}
\begin{figure}[h]
	\centering
	\begin{minipage}{0.45\textwidth}
		\includegraphics[width=0.85\textwidth]{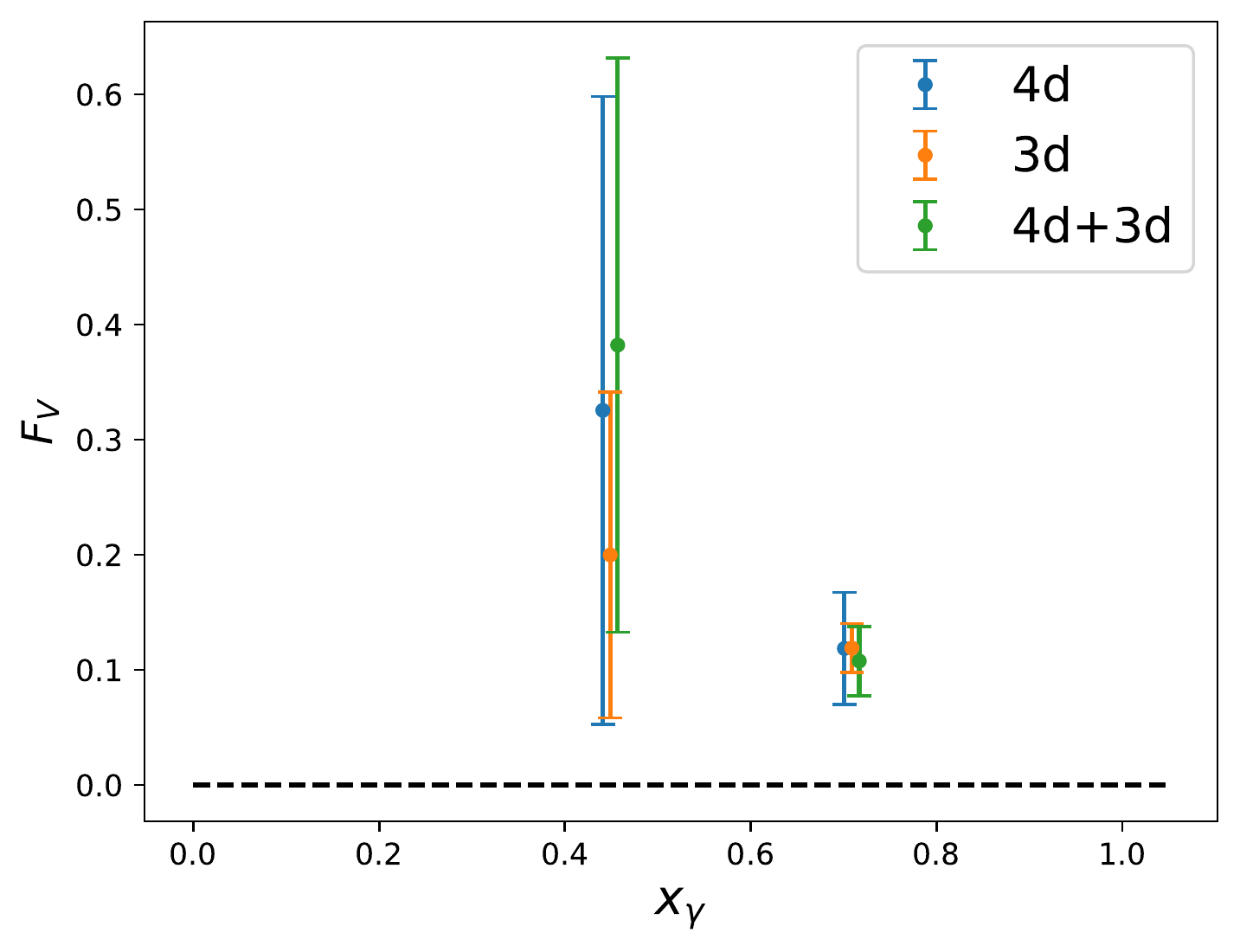}
	\end{minipage}
	\begin{minipage}{0.45\textwidth}
		\includegraphics[width=0.85\textwidth]{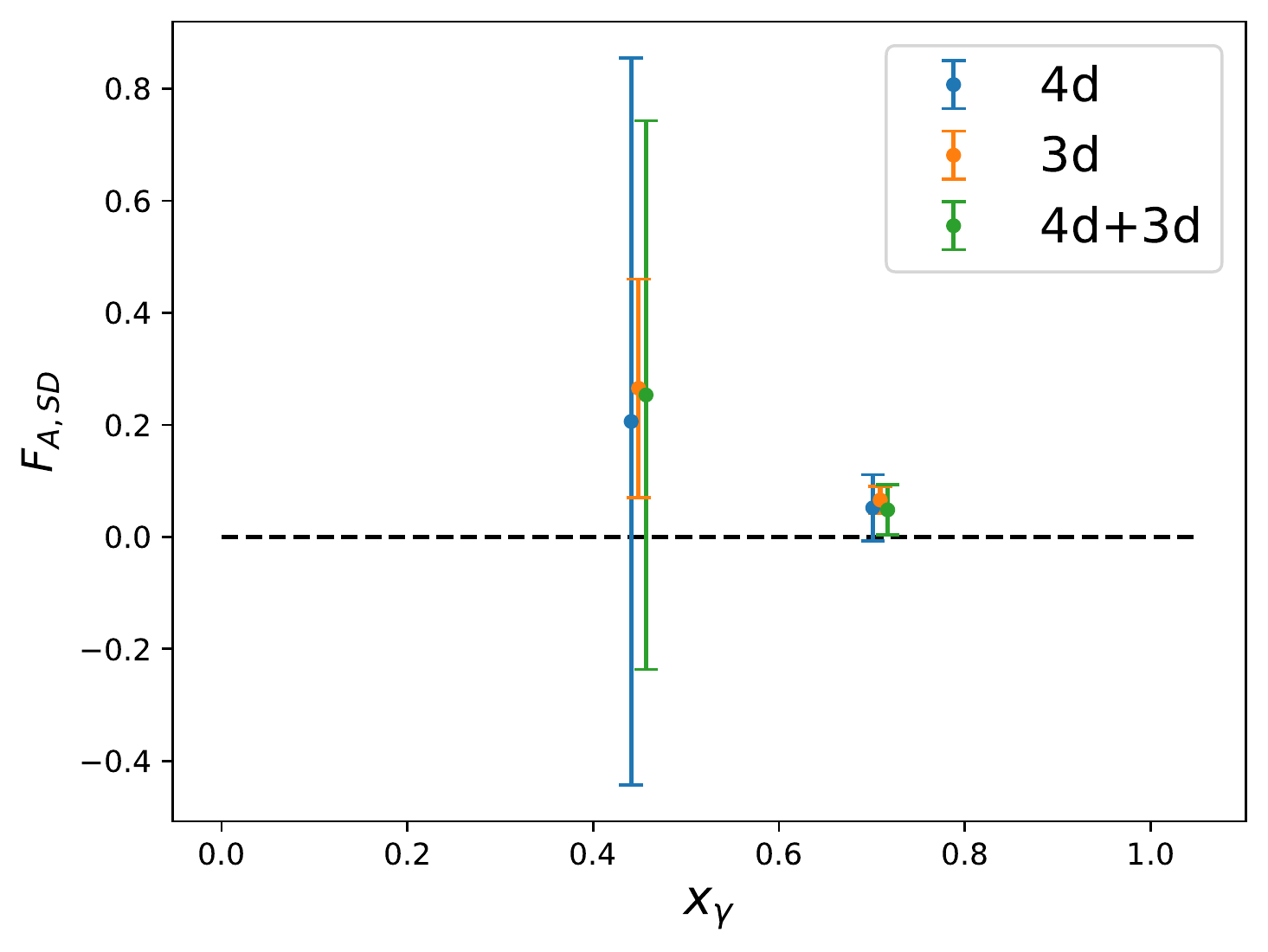}
	\end{minipage}
	\vspace{-0.15in}
	\begin{minipage}{0.45\textwidth}
		\includegraphics[width=0.85\textwidth]{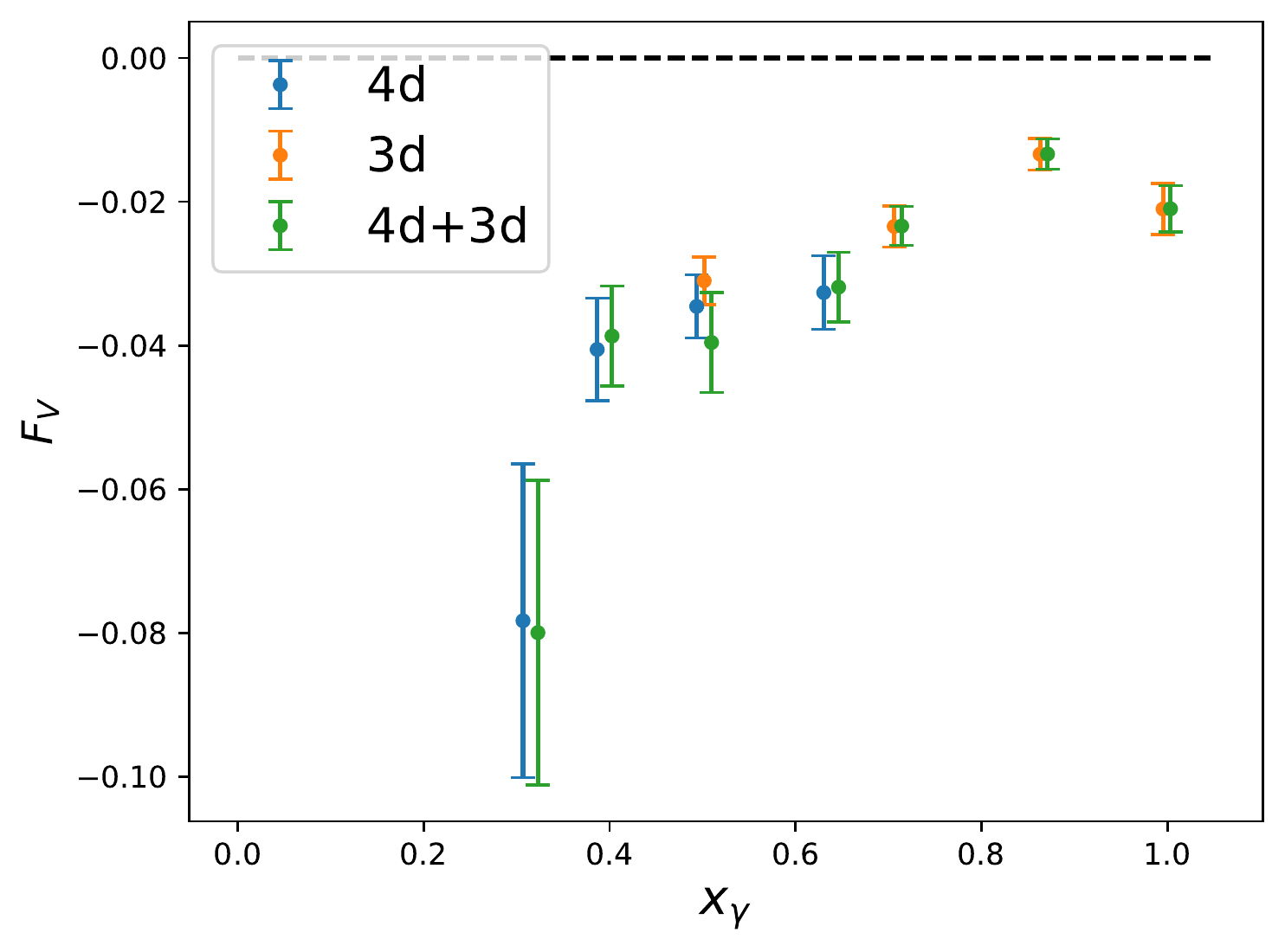}
	\end{minipage}
	\begin{minipage}{0.45\textwidth}
		\includegraphics[width=0.85\textwidth]{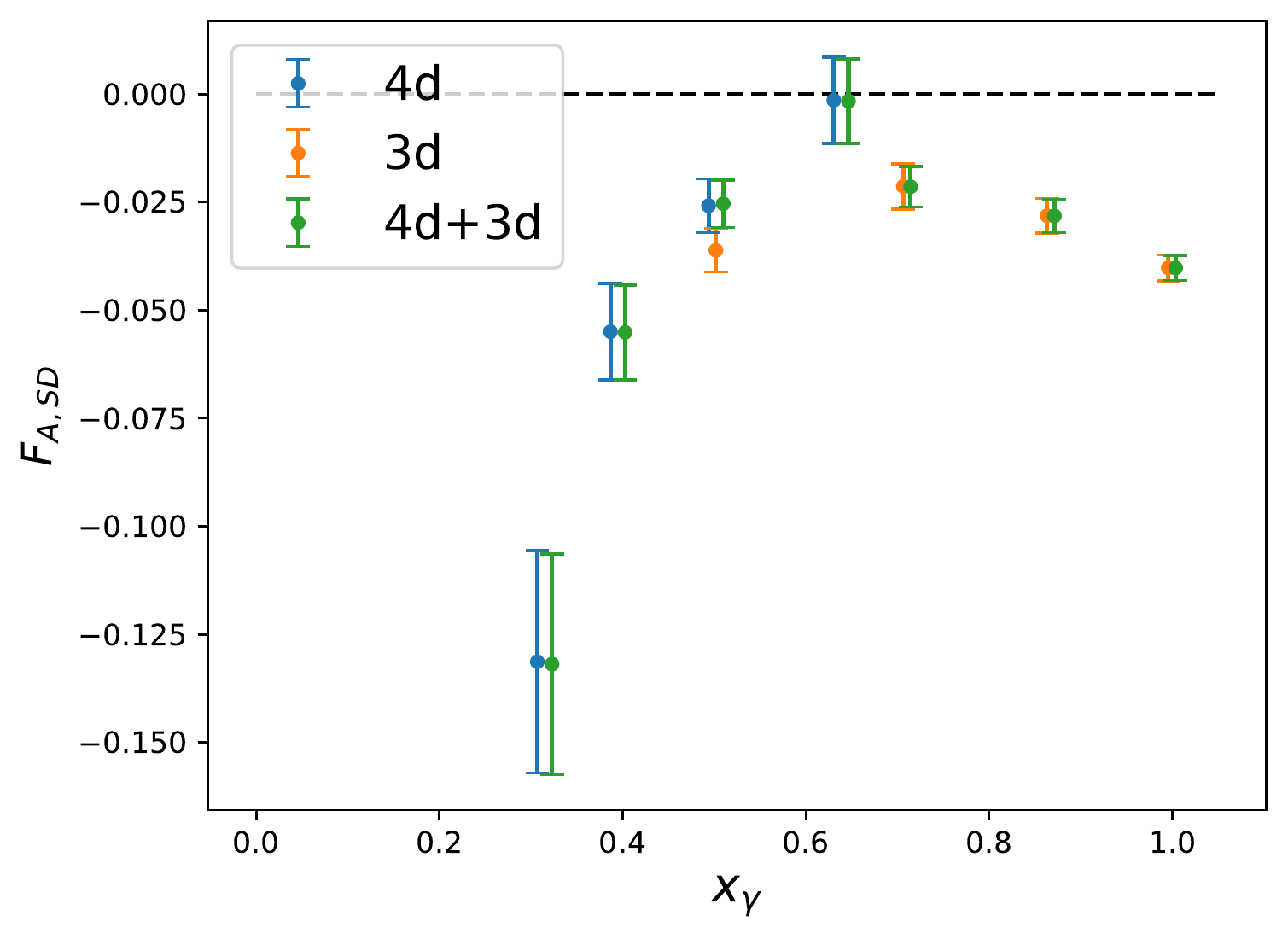}
	\end{minipage}
	\caption{The left (right) column show the global fit results for $F_V$ ($F_{A,SD}$) as a function of $x_\gamma$. The top (bottom) row shows $K^-$ ($D_s^+$) form factors. Different colored points show different combinations of data in the global fit. Data points at the same $x_\gamma$ value have been shifted slightly for clarity.}
	\label{fig:form_factors_vs_xgamma_method_comparison}
\end{figure}

The top plots in figure \ref{fig:form_factors_vs_xgamma_method_comparison} show the $K^- \to \ell^- \bar{\nu} \gamma$ form factors $F_V$ and $F_{A,SD}$ as a function of $x_\gamma$ calculated using 3d method data, 4d method data, and a combined analysis to both sets of data. First, we notice that the error bars are significantly larger for the 4d method fits compared to the 3d method fits. This is likely because the 4d method cannot resolve the sum of the unwanted exponentials of the separate time orderings. The combined analysis using both 3d and 4d method data allows us to remove the prior on the $E^>$ parameter that was necessary to stabilize the 4d method fits. We still use the prior on the excited-state energy gap from the two-point function. The combined global fit has approximately the same or increased statistical uncertainties when compared to the either the 3d or 4d method, which needs to be better understood. The bottom plots in figure \ref{fig:form_factors_vs_xgamma_method_comparison} show similar plots but for the $D_s^+ \to \ell^+ \nu \gamma$ form factors. The qualitative behavior of the fit results are similar to the Kaon decay. 

Using periodic boundary conditions, it is necessary for our current lattice sizes to perform the calculation in the moving frame of the Kaon to get physically allowed values of $x_\gamma$. Giving the pseudoscalar meson momentum increases the statistical noise significantly compared to the rest-frame calculations. The calculation in \cite{Desiderio:2020oej} was done using twisted boundary conditions, which has the advantage that small values of $x_\gamma$ can be achieved giving less momentum to the meson, improving statistical precision at small $x_\gamma$. However, in \cite{Desiderio:2020oej}, the maximum value of $x_\gamma$ obtained for the $D_s$ is $\sim 0.35$. Consequently, it was not possible to distinguish the $E_{\gamma}^{(0)}$ dependence of the form factors between a pole and polynomial form. In contrast, in the rest frame of the $D_s$ we obtain $x_\gamma$ up to the maximum allowed value of $x_\gamma=1$ and as low as $x_\gamma=0.5$. Using moving frames we have values as low as $x_\gamma=0.3$.
\vspace{-0.15in}
\section{Conclusion and future plans}
We have found that for certain values of $x_\gamma$, the fit results for the form factors do not plateau as we increase the integration range, and more complicated fits must be performed to remove the unwanted exponentials from intermediate states. We have calculated $I_{\mu \nu}(t_H, T)$ using a 3d and 4d sequential propagator, and compared analysis methods using only 3d data, only 4d data, and a combination of the two. Our comparison shows that the 3d method results in the smallest statistical uncertainties. However, because we only have 3 value of $t_H$, it is difficult to demonstrate stability in the $t_H$ fit ranges.

Moving forward, we will calculate the different time orderings of $I_{\mu \nu}(t_H, T)$ separately using the 4d method, which is expected to reduce the statistical uncertainty from the 4d method fits. Additionally, we will perform calculations using twisted boundary conditions to reach smaller $E_\gamma^{(0)}$ while giving less overall momentum to the meson. This is particularly important for the Kaon decay. Once the optimal analysis method has been worked out, we plan to perform calculations on a variety of ensembles and perform continuum and physical-pion-mass extrapolations for the $K^-$ and $D_s^+$ decays.
\vspace{-0.1in}
\subsubsection*{Acknowledgements:} We thank the RBC and UKQCD Collaborations for providing the gauge-field configurations. C.K. is supported by the DOE Computational Science Graduate Fellowship under award number DE-SC0020347. C.L. is supported in part by US DOE Contract DESC0012704(BNL). S.M. is supported by the U.S Department of Energy, Office of Science, Office of High Energy Physics under Award Number DE-SC0009913. A.S is supported in part by the U.S. DOE contract \#DE-SC0012704. We performed calculations using the QLUA software. We acknowledge NSF XSEDE, DOE Office of Science, and PRACE for awarding us access to TACC, NERSC, and GCS@LRZ, respectively.


\vspace{-0.1in}

\providecommand{\href}[2]{#2}\begingroup\raggedright\endgroup

\end{document}